\documentclass[journal,web,twoside,hidelinks]{ieeecolor}
\usepackage{generic}

\usepackage[english]{babel}
\usepackage[utf8x]{inputenc}
\usepackage[T1]{fontenc}
\usepackage{bm}
\usepackage{amssymb}
\usepackage{siunitx}

\newcommand{\refappendix}[1]{\hyperref[#1]{Appendix~\ref*{#1}}}
\newcommand{\refsupinf}[1]{\hyperref[#1]{Supplementary Information \ref*{#1}}}
\usepackage{cite}
\usepackage{amsmath}
\usepackage{empheq}
\usepackage[most]{tcolorbox}
\usepackage{graphicx}
\usepackage{epstopdf}
\AppendGraphicsExtensions{.gif}
\usepackage{amsfonts}
\usepackage{algorithmic}
\usepackage{textcomp}
\usepackage{todonotes}
\usepackage[colorlinks=false,linkcolor=.,hidelinks]{hyperref}
\usepackage{colortbl}
\usepackage[caption=off]{subfig}

\begin{document}
%% Math
\newcommand{\bv}[1]{\mathbf{#1}}
\newcommand{\Grad}{\mbox{\boldmath $\nabla$}}
\newcommand{\Div}{\Grad\!\cdot}
\newcommand{\Lap}{\mbox{\boldmath $\Delta$}}
\newcommand{\Curl}{\Grad\!\times}
\newcommand{\delsq}{\nabla^2}
\newcommand{\mean}[1]{\left\langle{#1}\right\rangle}
\newcommand{\infd}{\textrm{d}}
\newcommand{\diff}[2]{\frac{\infd #1}{\infd #2}}
\newcommand{\difftwo}[2]{\frac{\infd^2 {#1}}{\infd {#2}^2}}
\newcommand{\pdiff}[2]{\frac{\partial #1}{\partial #2}}
\newcommand{\pdiffcon}[3]{\left(\frac{\partial #1}{\partial #2}\right)_{#3}}
\newcommand{\pdifftwo}[2]{\frac{\partial^2 {#1}}{\partial {#2}^2}}
\newcommand{\pdifftwomix}[3]{\frac{\partial^2 {#1}}{\partial {#2}\,\partial {#3}}}
\newcommand{\vel}{\bv{v}}
\newcommand{\expu}[1]{\textrm{e}^{#1}}
\newcommand{\ehat}{\bv{e}}
\newcommand{\intomega}{ \mathop{}\!\int\limits_{\Omega}}
\newcommand{\intelement}{ \mathop{}\!\int\limits_{e_k}}
\newcommand{\intelementgen}{ \mathop{}\!\int\limits_{e_k^{xy}}}
\newcommand{\intelementref}{ \mathop{}\!\int\limits_{e_k^{\xi \eta}}}
\newcommand{\be}{\begin{equation}}
\newcommand{\ee}{\end{equation}}
\newcommand{\baa}{\begin{alignat}{2}}
\newcommand{\eaa}{\end{alignat}}
\newcommand\smathcal[1]{
  \mathchoice
    {{\scriptstyle\mathcal{#1}}}% \displaystyle
    {{\scriptstyle\mathcal{#1}}}% \textstyle
    {{\scriptscriptstyle\mathcal{#1}}}% \scriptstyle
    {\scalebox{.7}{$\scriptscriptstyle\mathcal{#1}$}}%\scriptscriptstyle
  }
\newcommand\sbmathcal[1]{
  \mathchoice
    {{\scriptstyle{\bm{\mathcal{#1}}}}}% \displaystyle
    {{\scriptstyle{\bm{\mathcal{#1}}}}}% \textstyle
    {{\scriptscriptstyle{\bm{\mathcal{#1}}}}}% \scriptstyle
    {\scalebox{.7}{$\scriptscriptstyle{\bm{\mathcal{#1}}}$}}%\scriptscriptstyle
  }

\newcommand{\note}[2]{ {\bf #1: }#2} 

\renewcommand{\sectionautorefname}{Section}
\renewcommand{\figureautorefname}{Fig.}% PS
\let\subsectionautorefname\sectionautorefname
\let\subsubsectionautorefname\sectionautorefname
\definecolor{lightgray}{rgb}{0.925,0.925,0.925}
\def\journalname{}
\def\titlename{Model-based reconstruction of non-rigid 3D motion-fields from minimal k-space data: MR-MOTUS}
\title{\titlename}
\author{Niek R. F. Huttinga, Cornelis A. T. van den Berg, 
Peter R. Luijten,
Alessandro Sbrizzi \thanks{This work was supported in part by the Netherlands Organisation for Scientific Research (NWO) under Grant 15115.} \thanks{The authors are with the Imaging Division of the University Medical Center, Utrecht,
Heidelberglaan 100, 3584 CX, Utrecht, The Netherlands. Correspondence should be directed to N.R.F. Huttinga (e-mail: n.r.f.huttinga@umcutrecht.nl)} \thanks{This work has been submitted to the IEEE for possible publication. Copyright may be transferred without notice, after which this version may no longer be accessible.} }

\def\BibTeX{{\rm B\kern-.05em{\sc i\kern-.025em b}\kern-.08em
    T\kern-.1667em\lower.7ex\hbox{E}\kern-.125emX}}
\markboth{\journalname}
{Huttinga \MakeLowercase{\textit{et al.}}: \titlename}

\maketitle
\begin{abstract}
Estimation of internal body motion with high spatio-temporal resolution can greatly benefit MR-guided radiotherapy/interventions and cardiac imaging, but remains a challenge to date. In image-based methods, where motion is indirectly estimated by reconstructing and co-registering images, a trade off between spatial and temporal resolution of the motion-fields has to be made due to the image reconstruction step. However, we observe that motion-fields are very compressible due to the spatial correlation of internal body motion. Therefore, reconstructing only motion-fields directly from surrogate signals or k-space data without the need for image reconstruction should require few data, and could eventually result in high spatio-temporal resolution motion-fields. In this work we introduce MR-MOTUS, a framework that makes exactly this possible. The two main innovations of this work are (1) a signal model that explicitly relates the k-space signal of a deforming object to general non-rigid motion-fields, and (2) model-based reconstruction of motion-fields directly from highly undersampled k-space data by solving the corresponding inverse problem. The signal model is derived by modeling a deforming object as a static reference object warped by dynamic motion-fields, such that the dynamic signal is given explicitly in terms of motion-fields. We validate the signal model through numerical experiments with an analytical phantom, and reconstruct motion-fields from retrospectively undersampled in-vivo data. Results show that the reconstruction quality is comparable to state-of-the-art image registration for undersampling factors as high as 63 for 3D non-rigid respiratory motion and as high as 474 for 3D rigid head motion. 
\newline\newline

\end{abstract}

\section{Introduction}

\IEEEPARstart{T}he estimation of motion-fields with high spatio-temporal resolution can greatly benefit applications such as MR-guided interventions/radiotherapy and cardiac imaging, but remains a challenge for the current state-of-the-art dynamic MRI methods. Previously proposed motion estimation methods in MRI can broadly be subdivided into three categories based on the required input data: images, surrogate signals, or $k$-space data. These categories will be briefly reviewed here. 

Image-based methods estimate motion from MRI data by reconstructing and co-registering images. For applications where time-resolved motion information is required, a trade off has to be made between the temporal and spatial resolution of the reconstructed images, and thus motion-fields, due to the inherently slow data-encoding rate of MRI. Parallel imaging (PI) \cite{pruessmann1999sense,griswold2002generalized} and compressed sensing (CS) \cite{lustig2008compressed} techniques have been proposed that reduce the required amount of data for the image reconstruction by exploiting coil sensitivity information and the compressibility of images \cite{jung2009k,otazo2015low}, but the achievable acceleration is still insufficient for the aforementioned high frame rate 3D applications. Alternatively, the temporal and spatial resolution can be decoupled by synchronizing data acquisition with the respiratory/cardiac cycle either prospectively through gating (see e.g. \cite{uribe2007whole}), or retrospectively through sorting (see e.g. \cite{feng2016xd}). However, both techniques only allow retrospective reconstructions, and are therefore not suitable for online applications like MR-guided radiotherapy. In addition, both gating and sorting assume periodic motion, which is not necessarily valid for cardiac motion, e.g. in arrhythmic patients, and may not be valid for respiratory motion due to hysteresis and varying end-inhale or end-exhale positions \cite{mcclelland2013respiratory}. Partial image-based methods have been proposed that resolve the trade off between temporal and spatial resolution by inferring 3D motion from partial or corrupted 3D image data. An example of this is the work by Lee \cite{lee2018robust}, where motion was estimated by co-registering severely corrupted images that were reconstructed from highly undersampled $k$-space data. Promising results were presented for rigid motion, but the application to non-rigid motion has yet to be demonstrated and will be challenging as the undersampling artefacts can yield reconstructions of unrealistic motion-fields. Another example of a partial image-based method is the work by Stemkens et al. \cite{stemkens2016image}, where 3D motion was inferred from orthogonal 2D time-resolved MR images through the registration of a 3D reference image to the 2D+time (2D+t) sequence using a low-dimensional PCA-based 3D motion model. 

A different category of motion estimation methods aims to reconstruct motion-fields from surrogate signals such as a time series from a respiratory belt \cite{mcclelland2013respiratory,mcclelland2017generalized} or the time evolution of noise covariances of an RF coil-array \cite{andreychenko2017thermal}. First, a correspondence model is fitted on training data that maps the surrogate signals to motion-fields. This model is then used to prospectively estimate motion from only the surrogate signal. A drawback of this type of method is that the surrogate signals are usually of poor quality and require additional hardware to be acquired. Besides that, the methods assume a very low-dimensional motion model, which either oversimplifies the reconstructed motion-fields or makes strong assumptions on the similarity with the training set.

Similarly to the surrogate signal methods, $k$-space methods are not limited by the image reconstruction step required in image-based methods. Several methods have been proposed to estimate motion directly from highly undersampled $k$-space data (see e.g. \cite{fu1995orbital,van2006real,welch2002spherical,pipe1999motion}). These methods are mainly based on the explicit relation between a linear transformation and $k$-space data due to properties of the Fourier transform \cite{wisetphanichkij2005fast} and are therefore limited to affine motion. Also joint image and motion reconstruction has been proposed (see e.g. \cite{burger2018variational,zhao2018coupling,odille2008generalized}), where alternatingly images and motion-fields are reconstructed directly from $k$-space data by inverting a coupled signal model. Here the dynamic sequence of reconstructed images is constraint to be consistent with a-priori knowledge on motion, and can therefore yield a more realistic solution. This approach has received much attention lately and has shown promising results, but it still requires the reconstruction of images which limits the achievable level of acceleration.

The aforementioned methods to estimate motion from MRI data all have drawbacks that may limit the practical application to MR-guided radiotherapy and MR-guided interventions: they either require image reconstructions which forces a trade off between temporal and spatial resolution, are limited to affine motion, or require additional hardware to acquire a surrogate signal.

In this work, we introduce a framework for Model-based Reconstruction of MOTion fields from Undersampled Signals (MR-MOTUS). The MR-MOTUS framework allows to reconstruct general non-rigid 3D motion-fields directly from $k$-space, without the requirement to reconstruct images. Internal body motion exhibits strong spatial correlations due to the connected mechanical structure of tissue, i.e. nearby tissue will likely move very similarly. We therefore hypothesize that motion-fields are very compressible, thus a very high level of acceleration can be achieved by reconstructing {\it only} motion-fields directly from the data. The two main innovations of this work that form the backbone of MR-MOTUS are (1) a dynamic MR-signal model that explicitly relates $k$-space data of a deforming object to general non-rigid motion-fields, and (2) solving the corresponding ill-posed non-linear inverse problem to reconstruct motion-fields from minimal data. We exploit the spatial compressibility of the motion-fields by using a lower-dimensional representation basis, which effectively reduces the dimension of the solution space. We validate the signal model through numerical experiments with an analytical phantom, and reconstruct motion-fields from retrospectively undersampled {\it in-vivo} head and {\it in-vivo} abdomen data. Results show that the reconstruction quality is comparable to state-of-the-art optical flow for undersampling factors as high as 63 for 3D non-rigid respiratory motion and as high as 474 for 3D rigid head motion.

% \subsection{Paper outline}
% This paper is organized as follows. In \autoref{section:ansatz} we illustrate the compressibility of motion-fields, and in \autoref{section:modelderivation} we derive the forward signal model that explicitly relates motion-fields to $k$-space data. In \autoref{section:problemformulation} we formulate the inverse problem that needs to be solved in order to reconstruct the motion-fields. The derived signal model is validated in \autoref{section:modelvalidation} by means of in-silico experiments with a deforming analytical phantom for which ground-truth motion-fields are available. Details on the different steps in the reconstruction process are presented in \autoref{section:theoryreconstruction}, and \autoref{section:methodsmotionestimation} describes the experimental setup that was used for several motion reconstruction experiments. Finally, reconstructions of 3D simulated phantom motion, as well as {\it in-vivo} head, and {\it in-vivo} respiratory motion are presented in \autoref{section:results}. Additionally, we show in \autoref{section:results} quantitative comparisons with the ground-truth and optical flow motion-fields for the in-silico experiment and {\it in-vivo} experiments, respectively.

\section{Theory}
\subsection{Ansatz}
\label{section:ansatz}
\begin{figure*}
    \centering
    \includegraphics[width=0.75\textwidth]{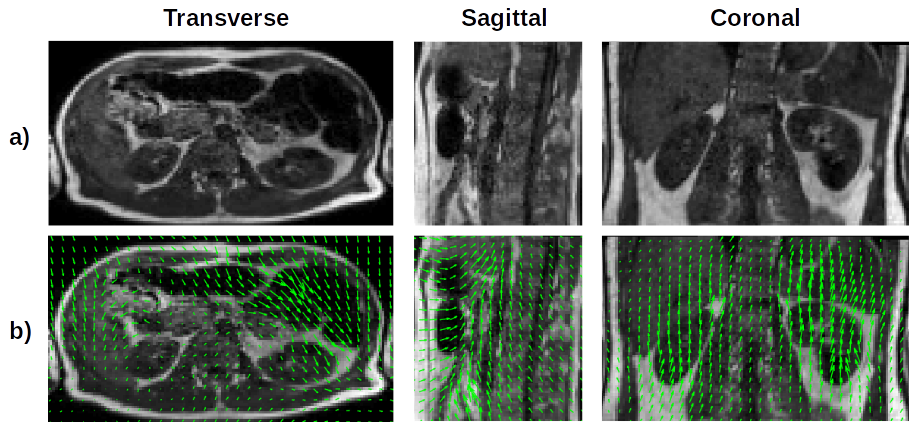}
    \caption{{\it This figure is best viewed online.} Visualization of the data used in \autoref{section:ansatz}: a) Three slices of one of the images used for the registration, and b) three in-plane projections of the motion-field obtained with optical flow.}
    \label{fig:compressiondata}
\end{figure*}

We illustrate the high compressibility of motion-fields by approximating them with a gradually decreasing number of basis functions from a natural representation basis. Two 3D abdomen scans were acquired during breath-holds in different respiratory phases using a spoiled gradient echo sequence with $\text{TR/TE}=2.30/1.15$ms, a FOV of $0.28 \times 0.34 \times 0.34$m and a resolution of $3.0 \times 2.7 \times 2.7$mm. To obtain a motion-field, the two images were registered using state-of-the-art optical flow software \cite{zachiu2015framework,zachiu2015improved}. One of the images and the obtained motion-field are shown in \autoref{fig:compressiondata}.

\begin{figure}[tbp]
\centering    
\includegraphics[width=0.49\textwidth]{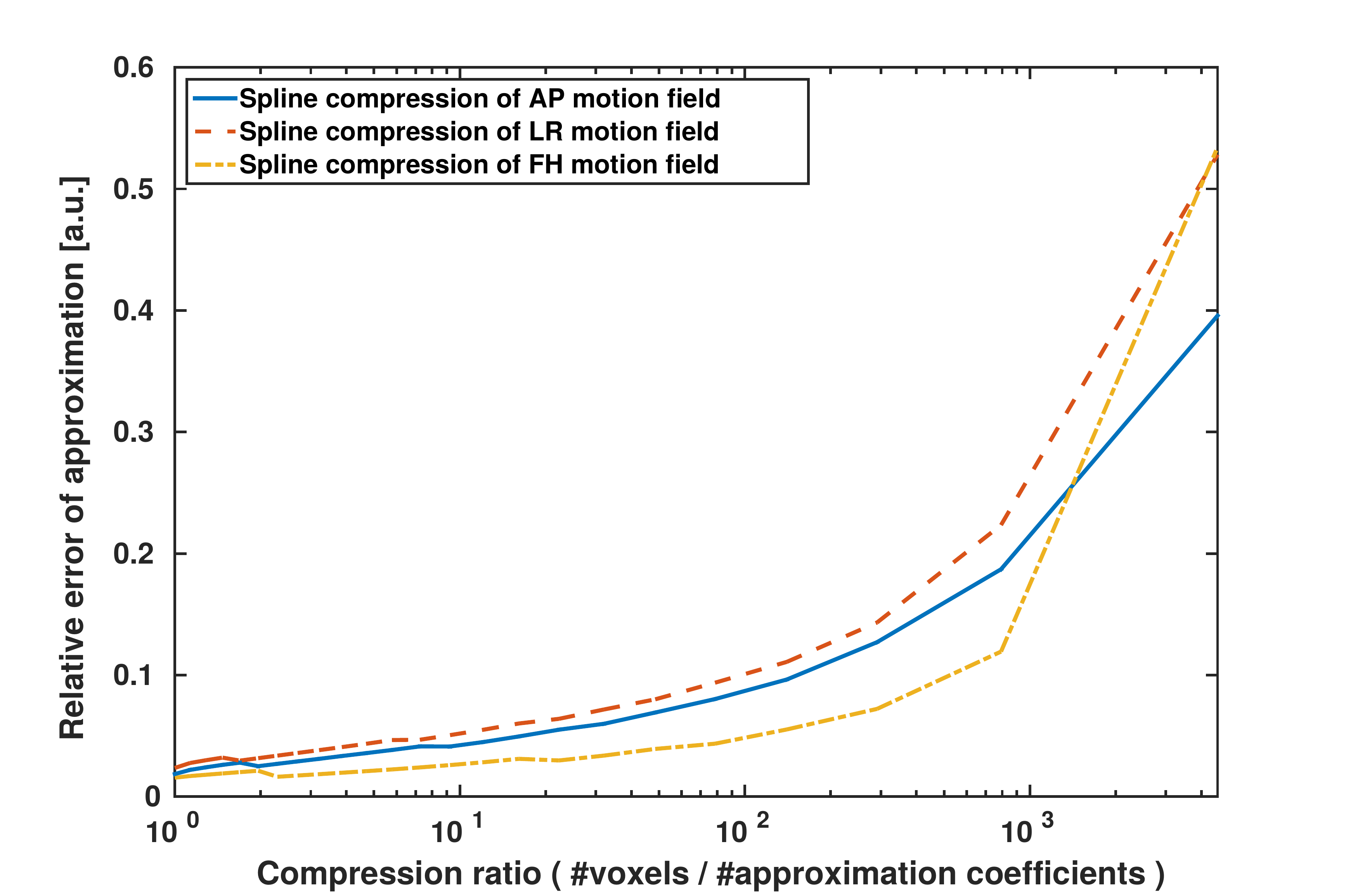}    
\caption{Compression ratios for the three components of the motion-field of respiratory motion in a cubic B-spline basis, as described in \autoref{section:ansatz}. The curves show that the motion-fields are indeed very compressible: a maximum representation error of only 10\% is made for all three components with 100 times as little approximation coefficients.}
\label{fig:compressioncurves}
\end{figure}

Next, a cubic B-spline basis \cite{rueckert1999nonrigid} was chosen as the natural representation basis and all components (LR, FH, AP) of the motion-field were represented separately. The maximum relative error\footnote{In this work the relative error $e$ between a vector $\bv{a}$ and a target vector $\bv{b}$ is defined as $e = 
\frac{\lVert \bv{a} - \bv{b} \rVert}{\lVert \bv{b} \rVert}$.} of approximation over all three components (LR, FH, AP) was computed at several compression ratios. Here we have defined the compression ratio as the ratio between the number of voxels in the motion-field and the number of basis functions. The compression curves for all components are shown in \autoref{fig:compressioncurves}. Note that the LR component gave the highest representation error at all compression ratios. A maximum representation error of only 10\% is made for all three components with 100 times as few approximation coefficients, which shows that the motion-fields are indeed very compressible.

\subsection{Signal model derivation}
\label{section:modelderivation}
\subsubsection{Outline of the derivation}
Let $q_t(\bv{r})\in \mathbb{C}$ denote the transverse magnetization of a deforming object at time $t$ and spatial coordinate $\bv{r}=(x,y,z)$.
The $k$-space signal from $q_t$ at coordinate $\bv{k}=(k_x,k_y,k_z)$ can then be modeled as
\begin{equation} \label{eq:mrsignal2}
    s_t(\bv{k})=\int_{\Omega} q_t(\bv{r}) e^{-i 2 \pi \bv{k} \cdot \bv{r}} \ \infd \bv{r}.
\end{equation}
Here $\Omega$ denotes the spatially excited FOV. Let $\bv{U}_t: \mathbb{R}^3 \mapsto \mathbb{R}^3$ denote the motion-field that deforms $q_0$ to $q_t$:
\begin{equation}
\bv{U}_t(\bv{r}) = \bv{r} + \bm{\delta}_t(\bv{r}),
\end{equation}
with displacement function $\bm{\delta}_t: \mathbb{R}^3 \mapsto \mathbb{R}^3$. Under several assumptions, that we will later elaborate upon, the time-dependent local signal contribution can be written as the deformation of $q_0$ by motion-fields $\bv{U}_t$:
\begin{equation}
    \label{eq:mainsubstitutionq}
    q_t(\bv{r})\infd \bv{r} = q_0\left(\bv{U}_t(\bv{r})\right)\lvert \det(\Grad \bv{U}_t) (\bv{r})\rvert \infd \bv{r}.
\end{equation}
Here $\Grad \bv{U}_t$ denotes the Jacobian of the motion-field $\bv{U}_t$. In the rest of this work we will refer to $q_0$ as the {\it reference image} and to $s_t$ as the {\it snapshot signal}. An explicit relation between the reference image, motion-fields, and the snapshot signal of the deforming object can be obtained by substituting \eqref{eq:mainsubstitutionq} into \eqref{eq:mrsignal2}, followed by a change of coordinates:
\begin{equation}
\label{eq:forwardmodelpreview}
s_t(\bv{k})=\int_{\Omega} q_0(\bv{r}_0) e^{-i 2 \pi \bv{k} \cdot \bv{T}_t(\bv{r}_0)} \ \infd \bv{r}_0.
\end{equation}
Here $\bv{T}_t$ is defined as the right inverse of $\bv{U}_t$, such that 
\begin{equation}
\label{eq:definitioninversedvf}
\bv{T}_t(\bv{r})=\bv{r} + \bm{\eta}_t(\bv{r}), \quad \bv{U}_t \circ \bv{T}_t = \text{\bfseries Id},
\end{equation}
where $\bm{\eta}_t(\bv{r})$ is the displacement of $\bv{r}$ due to $\bv{T}_t$ and $\text{\bfseries Id}$ denotes the identity operator.

Equation \eqref{eq:forwardmodelpreview} forms the backbone of the presented framework, and will therefore be referred to as the MR-MOTUS signal model. The details of the derivation of this model will be shown in the subsequent sections, but the outline is as follows. We will first separate the transverse magnetization $q_t$ into the unit-length transverse magnetization $m_t:\mathbb{R}^3\mapsto\mathbb{C}$ and spin density $\rho_t:\mathbb{R}^3\mapsto\mathbb{R}^+$ as 
    \begin{equation}
        q_t \equiv m_t \cdot \rho_t.
    \end{equation}
Next, we will derive temporal relations for $m$ and $\rho$ separately and combine them to obtain \eqref{eq:mainsubstitutionq}. Finally, we derive the MR-MOTUS signal model \eqref{eq:forwardmodelpreview} by a substitution and a change of variables.

\subsubsection{Temporal relation for transverse magnetization}
We first derive the temporal relation for the transverse magnetization $m$. Suppose a steady-state sequence is employed to acquire signal from a static object deformed by dynamic motion-fields $\bv{U}_t$. We assume a sufficiently short read-out time, such that spin displacements and transverse and longitudinal decay effects have a noticeable effect only over one or several TR intervals. The transverse magnetization at time $t$ of the spins at location $\bv{r}_t$ can then be written as the transverse magnetization at time $0$ of the same spin before deformation by $\bv{U}_t$: 
\begin{flalign}
\label{eq:steadystate}
    & m_t(\bv{r}_t) = m_0\left(\bv{U}_t(\bv{r}_t)\right). \\
    &\text{\it (Steady-state condition)} \nonumber  
\end{flalign}
Note that for \eqref{eq:steadystate} to hold it must be assumed that the $B_0$ and $B_1$ fields are spatially slowly varying, which is a reasonable assumption at the targeted clinical field strength of 1.5 tesla. Formally, \eqref{eq:steadystate} only makes sense under a few other technical assumption, and we refer to Supplementary Information, Section A, at the end of this text for a mathematically more formal derivation.

\subsubsection{Temporal relation for spin density}
\label{section:spins}
Next, we derive the equation for the spin density. To be able to describe the complete dynamic behaviour of a deforming object in terms of a static reference image and dynamic motion-fields, it is assumed that mass is conserved within the FOV. Under this assumption, the total number of excited spins during every TR remains constant, that is $$\int_{\Omega} \rho_t(\bv{r}_t)\ \infd \bv{r}_t = C, \ \ \ t=0,1,\dots,$$ where $C\in\mathbb{R}$ is a constant. Hence, the deformations $\bv{U}_t$ must satisfy 
\begin{equation} 
\int_{\bv{U}_t(X)} \rho_0(\bv{r}_0) \ \infd \bv{r}_0 = \int_X \rho_t(\bv{r}_t) \ \infd \bv{r}_t,
\label{eq:mapdefinition}
\end{equation}
for all sets $X \subseteq \Omega$. We assume all $\bv{U}_t$ are continuously differentiable everywhere and right invertible, with right inverse $\bv{T}_t$ as defined in \eqref{eq:definitioninversedvf}.
We can then rewrite the left-hand side of \eqref{eq:mapdefinition} using the change-of-variables $\bv{r}_0 \mapsto \bv{U}_t(\bv{r}_t)$:
\begin{equation} \int_{\bv{U}_t(X)} \rho_0(\bv{r}_0)\ \infd \bv{r}_0 = \int_{X} \rho_0\left(\bv{U}_t(\bv{r}_t)\right) |\det(\Grad \bv{U}_t )(\bv{r}_t) | \ \infd \bv{r}_t.\label{eq:spinconservation}
\end{equation} 
Combining \eqref{eq:mapdefinition} and \eqref{eq:spinconservation} then yields
\begin{equation}
    \int_{X} \rho_t(\bv{r}_t) \ \infd \bv{r}_t = \int_{X} \rho_0\left(\bv{U}_t(\bv{r}_t)\right) \lvert\det(\Grad \bv{U}_t) (\bv{r}_t)\rvert \ \infd \bv{r}_t,
\end{equation}
for all sets $X \subseteq \Omega$. We conclude that the following must hold for all $t=0,1,\dots$ and $\bv{r} \in \Omega$:
\begin{flalign}
    \label{eq:acspinconservation}
    \qquad\qquad& \rho_t(\bv{r}_t)\infd \bv{r}_t = \rho_0\left(\bv{U}_t(\bv{r}_t)\right)\lvert \det(\Grad \bv{U}_t ) (\bv{r}_t)\rvert \infd \bv{r}_t . \\
    \qquad\qquad& \text{\it (Mass conservation)}  \hfil \nonumber
\end{flalign}
\subsubsection{Derivation of the MR-MOTUS signal model}
Combining \eqref{eq:steadystate} and \eqref{eq:acspinconservation} yields the previously described temporal relation \eqref{eq:mainsubstitutionq} between the reference object, deforming object and motion-fields: 
\begin{equation}
    \label{eq:mainsubstitutionq2}
    q_t(\bv{r})\infd \bv{r}= q_0\left(\bv{U}_t(\bv{r})\right)\lvert \det(\Grad \bv{U}_t) (\bv{r})\rvert \infd \bv{r}.
\end{equation}
Substituting \eqref{eq:mainsubstitutionq2} into the signal model \eqref{eq:mrsignal2} then yields
\begin{eqnarray*}
s_t(\bv{k}) &=& \int_{\Omega} q_0\left(\bv{U}_t(\bv{r}_t)\right) e^{-i 2 \pi \bv{k} \cdot \bv{r}_t} \ \lvert \det(\Grad \bv{U}_t ) (\bv{r}_t) \rvert \ \ \infd \bv{r}_t.
\end{eqnarray*}
By the inverse function theorem the determinant of the inverse is the inverse of the determinant, hence 
after the change of variables $\bv{r}_t \mapsto \bv{T}_t(\bv{r}_0)$ we obtain
\begin{alignat}{2}
\label{eq:signalmodel}
s_t(\bv{k}) &= \int_{\bv{U}_t(\Omega)} q_0(\bv{r}_0) e^{-i 2 \pi \bv{k} \cdot \bv{T}_t(\bv{r}_0)} \ \infd \bv{r}_0.
\end{alignat}
Note that here we have used the right inverse property of $\bv{T}_t$, i.e. $\bv{U}_t \circ \bv{T}_t = \text{\bfseries Id}$. Due to the mass conservation assumption, the integration domain in \eqref{eq:signalmodel} can be changed to the spatially excited FOV $\Omega$ to obtain the final signal model:
\begin{equation}
\label{eq:signalmodel_approx}
s_t(\bv{k})=\int_{\Omega} q_0(\bv{r}_0) e^{-i 2 \pi \bv{k} \cdot \bv{T}_t(\bv{r}_0)} \ \infd \bv{r}_0.
\end{equation}
The assumption on the conservation of mass during deformation may seem restrictive, but it is not too hard to guarantee in at least 2 dimensions with a 3D FOV. For a moving head, the FOV must be taken such that the whole head is covered with the FOV during acquisition of all snapshot signalsS. For respiratory motion the FOV can simply be set to the largest expected extend of the body contours in AP and LR directions and as large as possible in the FH direction.
\subsection{Inverse problem formulation}
\label{section:problemformulation}
Note that \eqref{eq:signalmodel_approx} explicitly relates the $k$-space signal to a reference object through (possibly non-rigid/non-linear) motion-fields $\bv{T}_t$. If a reference image $q_0$ and snapshot data $\bv{s}_t$ are available, then motion can be estimated by solving the inverse problem corresponding to \eqref{eq:signalmodel_approx}. In order to exploit the compressibility of motion-fields, we represent them in a lower-dimensional basis using coefficients $\bm{\theta}_t \in \mathbb{R}^{N_c}$, with $N_c \ll 3N$. Equation \eqref{eq:signalmodel_approx} can then be rewritten in operator form as
\begin{equation}
\label{eq:forwardmodel}
\bv{s}_t = \bv{F}(\bm{\theta}_t | q_0 ), 
\end{equation}
where $\bv{F}(\bm{\theta}_t | q_0 )$ is the vectorization over $k$-space coordinates of
\[F(\bm{\theta}_t | q_0)[\bv{k}] = \int_\Omega q_0(\bv{r}_0) e^{-i 2 \pi \bv{k} \cdot \bv{T}_t(\bv{r}_0 \hspace{.04cm} | \hspace{.04cm} \bm{\theta}_t )} \ \infd \bv{r}_0, \]
and $\bv{s}_t$ is the vectorized $k$-space signal of the deforming object at time $t$. In the rest of this work we drop the dependency of $\bv{F}$ on $q_0$ for ease of notation, and because it assumed to be known. To reconstruct motion-fields the following minimization problem is solved:
\begin{equation}
\label{eq:inverseproblem}
\min_{\bm{\theta}_t} \    \lVert \bv{F}( \bm{\theta}_t ) - \bv{s}_t \rVert_2^2 + \lambda \mathcal{R}(\bv{T}_t(\cdot  \hspace{.04cm} | \hspace{.04cm} \bm{\theta}_t)),
\end{equation}
where $\mathcal{R}$ is a regularizer that models a-priori knowledge on motion-fields, and $\lambda\in\mathbb{R}^+$ is the corresponding regularization coefficient that balances the objective function between a data-fit and being consistent with the a-priori assumptions.

We would like to stress again the benefit of the proposed method. Since we are not solving for the images $\bv{q}_t$ but for $\bm{\theta}_t$, the amount of data required for motion estimation will not depend on spatial resolution directly but only on the dimensionality of the motion model parameterization through $\bm{\theta}_t$. Hence, when proper regularization is applied and a proper motion model is chosen, solving for $\bm{\theta}_t$ can be done with very fast snapshot acquisitions $\bv{s}_t$. In other words, the framework has the potential to estimate non-rigid motion-fields directly from minimal $k$-space data. 

In the next section the MR-MOTUS signal model \eqref{eq:signalmodel_approx} is validated through an in-silico experiment in which ground-truth motion-fields are available. Moreover, the details of solving \eqref{eq:inverseproblem} and obtaining $\bv{U}_t$ from $\bv{T}_t$ are discussed.

\section{Methods}

\subsection{Signal model validation}
\label{section:modelvalidation}
In order to validate the proposed signal model \eqref{eq:signalmodel}, experiments were performed on an analytical phantom with known ground-truth motion-fields. The analytical phantom was deformed with a pre-defined affine transform, and the signal from the deforming phantom was computed using \eqref{eq:mrsignal2} and compared with the signal obtained from the forward model \eqref{eq:signalmodel_approx}.

\begin{figure*}[tbp]
\centering
\subfloat{\includegraphics[width=.34\textwidth,height=0.23 \textwidth]{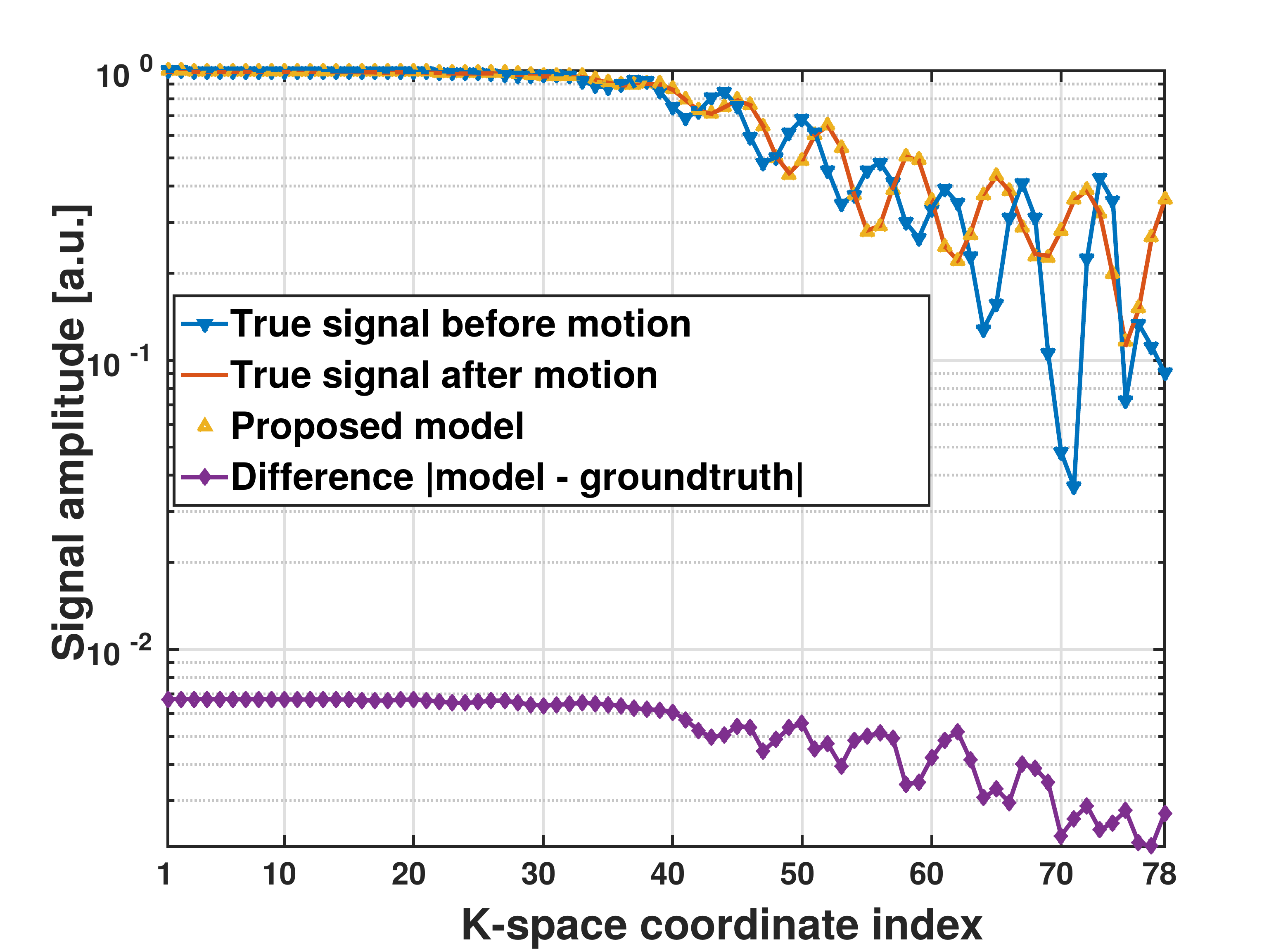}}
\subfloat{\includegraphics[width=.34\textwidth,height=0.23 \textwidth]{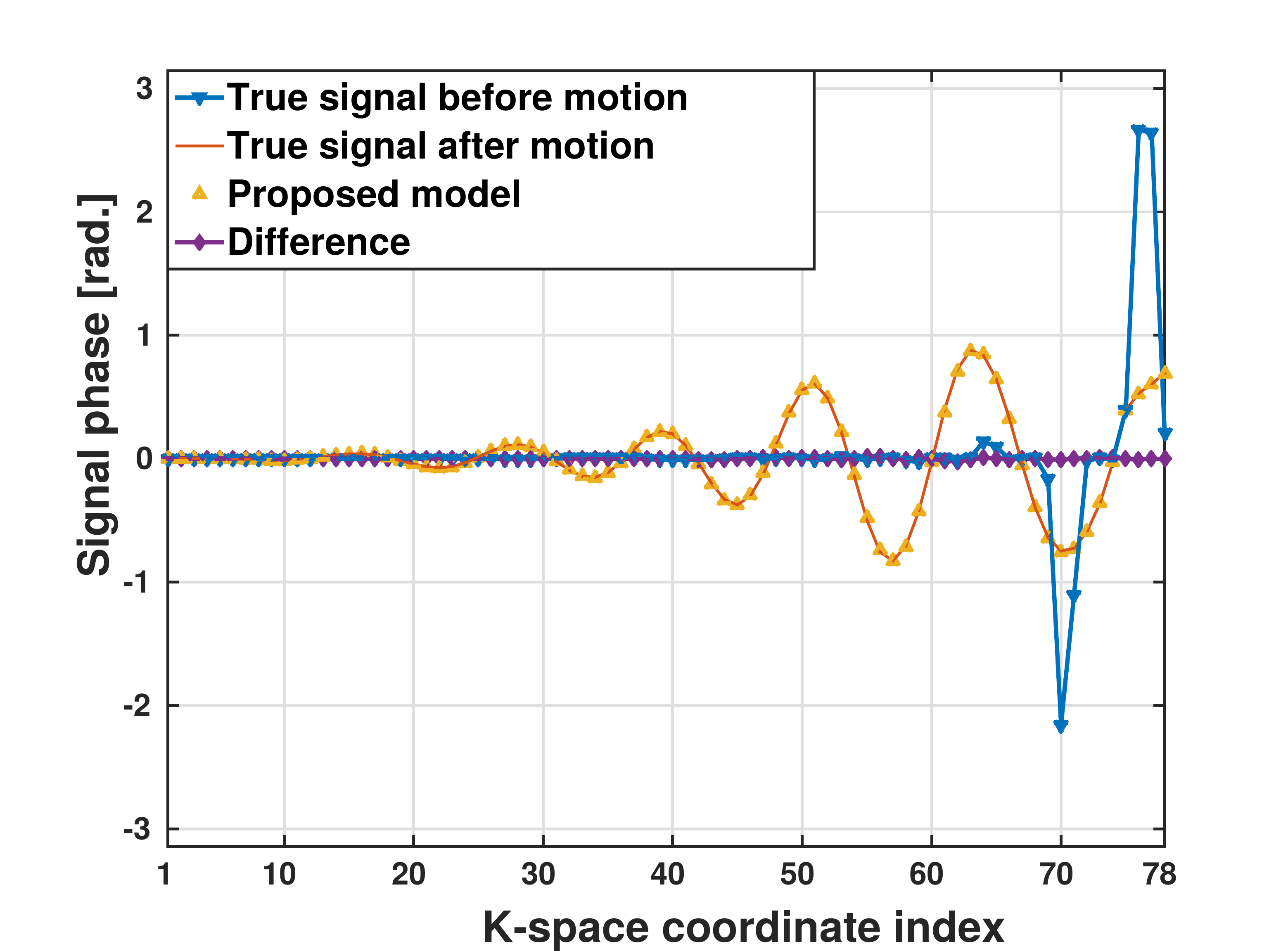}}
\subfloat{\includegraphics[width=0.27\textwidth,height=0.24 \textwidth]{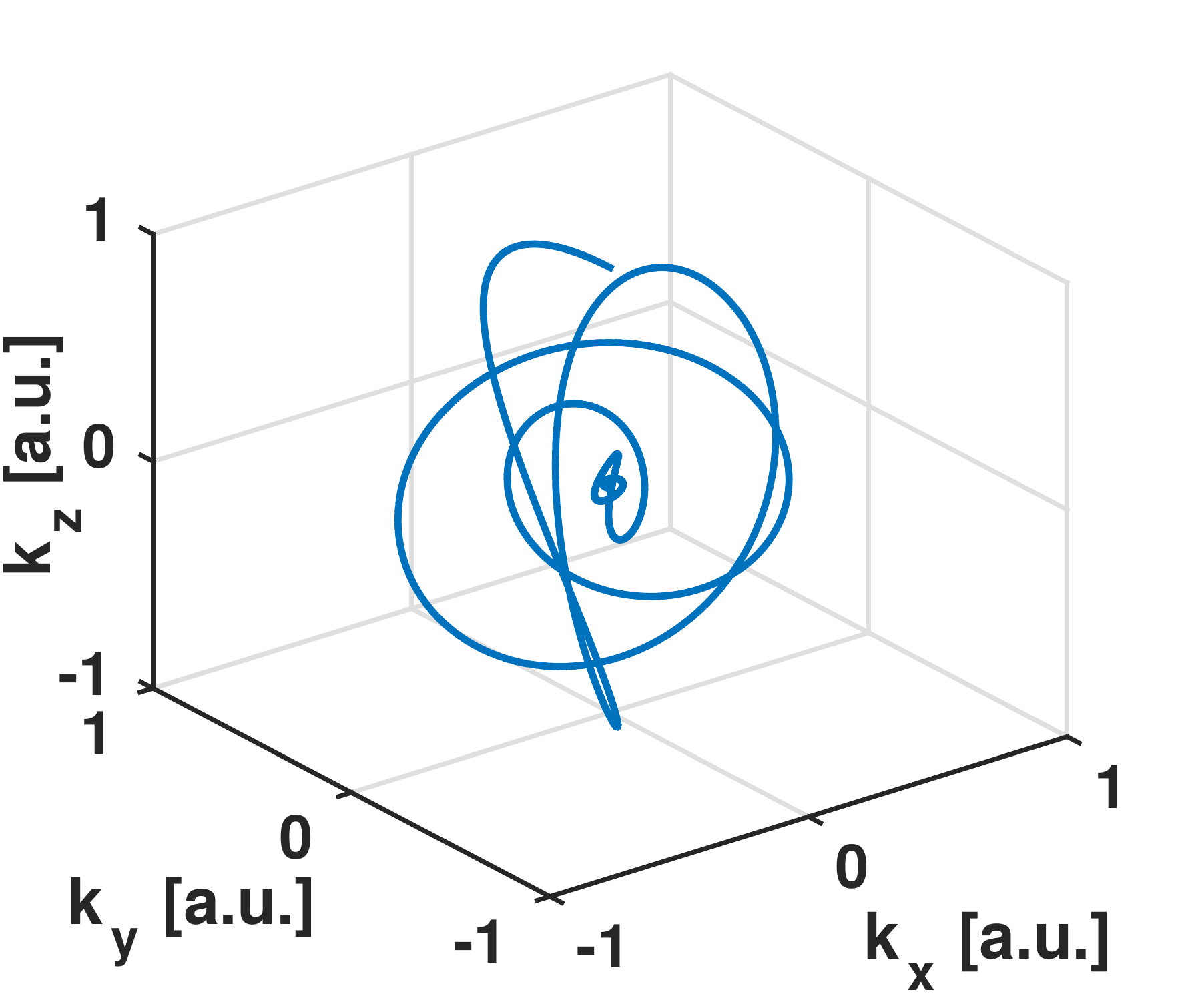}}
\caption{Results of the signal model validation in \autoref{section:modelvalidation}. Plots of the magnitude (left) and phase (middle) of the signal before deformation (blue), true signal after deformation with an affine transformation (red), the signal calculated with the proposed MR-MOTUS signal model \eqref{eq:signalmodel_approx} (yellow), and the difference between the model's signal and true signal (purple). The trajectory used for signal simulation is plotted on the right and was generated using the code provided by \cite{malik2012tailored}. 
Note that the model's signal and the true signal are indistinguishable; the difference is about two orders of magnitude lower than the signal. The small deviations between predictions by the model and ground-truth are likely caused by discretization errors.}
\label{fig:signalcomparison}
\end{figure*}

The phantom was modeled as the function $q_0^{\text{ph}}: \mathbb{R}^3 \mapsto \mathbb{R}$
\begin{alignat}{2}
\label{eq:analyticalphantom}
q_0^{\text{ph}}(x,y,z)=&e^{-\left(\frac{x^2}{0.15}+\frac{y^2}{0.08}+\frac{(z+0.20)^2}{0.10}\right)} + \nonumber \\  &0.85 \cdot e^{-\left(\frac{x^2}{0.15}+\frac{(y-0.25)^2}{0.08}+\frac{(z-0.25)^2}{0.10}\right)}.    
\end{alignat}
A grid of $128 \times 128 \times 128$ was used to discretize the phantom, and the spatial FOV was scaled to $[-1,1]$ in arbitrary units. The reference image $q_0^{\text{ph}}$ was deformed to a new object $q_1^{\text{ph}}$ with an affine transformation $\bv{U}_1$ using \eqref{eq:mainsubstitutionq2}:
\begin{equation}
\label{eq:acspinconservationphantom} 
q_1^{\text{ph}}(\bv{r}) = q_0^{\text{ph}}(\bv{U}_1(\bv{r})) \lvert \det \left(\bv{\Grad} \bv{U}_1  \right )(\bv{r}) \rvert. \end{equation} 
The affine transformation was constructed as the composition of a clockwise rotation by 45\textdegree \ around the vector $\begin{bmatrix} 0.9 & 0.1 & -0.3\end{bmatrix}^T$, a component-wise scaling with the vector $\begin{bmatrix}0.8 & 1.2 & 0.9\end{bmatrix}^T$, and a shift of $\begin{bmatrix}0.1 & -0.1 & 0.05\end{bmatrix}^T$. The true signal from the object before and after motion was computed using \eqref{eq:mrsignal2} at $t=0$ and $t=1$ respectively. The signal from our model \eqref{eq:signalmodel} was computed using only $q_0^{\text{ph}}$ and $\bv{T}_1$. Here $\bv{T}_1$ is the inverse of the affine transformation $\bv{U}_1$, which could easily be computed using linear algebra. A total number of $78$ $k$-space points were simulated using \eqref{eq:mrsignal2} with a very short (sub-ms), single-shot, 3D spiral trajectory (see \autoref{fig:signalcomparison}). The spiral was generated using the code provided by Malik et al. \cite{malik2012tailored}. The true signals before and after motion and the signal from our model are plotted in \autoref{fig:signalcomparison}. Note that the model's prediction of the signal on the spiral after deformation is indistinguishable from the true signal on the spiral after deformation. The minor deviations are likely caused by discretization of the continuous analytical phantom.
\subsection{MR-MOTUS: Model-based reconstruction of motion-fields from undersampled signals}
\label{section:theoryreconstruction}

\subsubsection{Regularization functional}
A natural choice for $\mathcal{R}$ in this setting, which was originally proposed in \cite{fischer2003curvature}, is to assume smooth motion-fields by penalizing the spatial curvature of the motion-fields: \[ \mathcal{R}(\bv{T}_t(\cdot  \hspace{.04cm} | \hspace{.04cm} \bm{\theta}_t)):=\sum_{p\in\{x,y,z\}} \int_\Omega  | \Delta T_t^p(\bv{r} \hspace{.04cm} | \hspace{.04cm} \bm{\theta}^p_t) |^2 \ \infd \bv{r}.\] Here $\Delta$ denotes the Laplace operator, and $T_t^x,T_t^y,T_t^z: \mathbb{R}^3 \mapsto \mathbb{R}$ denote the individual components of the motion-field $\bv{T}_t$. Since all linear transformations are contained in the kernel of the Laplace operator, this regularizer automatically performs the rigid alignment step that usually precedes the actual registration \cite{fischer2003curvature}. Using this smoothness prior we obtain the following minimization problem to reconstruct the motion model parameters $\bm{\theta}_t=\{\bm{\theta}_t^x, \bm{\theta}_t^y, \bm{\theta}_t^z\}$:
\begin{equation}
    \label{eq:inverseproblem2}
    \min_{\bm{\theta}_t} \    \lVert \bv{F}(\bm{\theta}_t)- \bv{s}_t \rVert_2^2 \ + \ \lambda \hspace{-.2cm} \sum_{p\in\{x,y,z\}} \int_\Omega | \Delta  T_t^p(\bv{r} \hspace{.04cm} | \hspace{.04cm} \bm{\theta}_t^p) |^2 \ \infd \bv{r}.
\end{equation}
\subsubsection{Motion models}
In this work two motion models are considered: 3D affine transformations and free-form deformations (FFD) parameterized using cubic B-splines \cite{rueckert1999nonrigid}. The 3D affine transformation is defined as
\begin{equation}
    \label{eq:affinemotionmodel}
    \bv{T}^{\text{aff}}(\bv{r} \hspace{.04cm} | \hspace{.04cm} \bv{A} , \bv{v}) = \bv{A}\bv{r} + \bv{v},
\end{equation}
where $\bv{A} \in \mathbb{R}^{3 \times 3}$ is the affine matrix and $\bv{v} \in \mathbb{R}^{3 \times 1}$ is the shift vector. This results in $3\times 3 + 3 = 12$ parameters. The free-form deformation is defined as
\begin{equation}
    \label{eq:bsplinemotionmodel}
\bv{T}^{\text{FFD}}(\bv{r} \hspace{.04cm} | \hspace{.04cm} \bv{c}^x,\bv{c}^y,\bv{c}^z) = \bv{r} + \begin{pmatrix} \bv{b}^x(\bv{r})\bv{c}^x \\ \bv{b}^y(\bv{r})\bv{c}^y \\ \bv{b}^z(\bv{r})\bv{c}^z \end{pmatrix},
\end{equation}
where $\bv{b}^p(\bv{r}) \in \mathbb{R}^{1 \times N_c^p}$ are the row-vectors with $N_c^p$ 3D B-spline basis functions, evaluated at the coordinate $\bv{r}$, and $\bv{c}^p \in \mathbb{R}^{N_c^p \times 1}$ denote expansion coefficients. The 3D basis functions are constructed as a Kronecker product of three 1D bases with a spline order (i.e. the number of basis functions) of $S$ each. If we use the same basis for all three components of the motion-field, then the total number of coefficients for the spline model is $N_c = \sum_p N_c^p = 3 S^3$. In practice this usually implies $N_c \approx \mathcal{O}(10^4)$ for a $100\times100\times100$ motion-field. In contrast, reconstruction of a $100 \times 100 \times 100$ image has a total number of unknowns in $\mathcal{O}(10^6)$, which is two orders of magnitude higher than the spline model.
\subsubsection{Optimization}
Solving the optimization problem in \eqref{eq:inverseproblem2} is challenging as it is both non-convex and non-linear. 
Nevertheless, various algorithms exist to tackle problems of this type. Most of these are based on Newton's method, where iterations of the form
\begin{eqnarray}
\label{eq:newtonsmethod}
\bm{\theta}^{(j+1)}=\bm{\theta}^{(j)}- \left[ \bv{H} \bv{F}\left(\bm{\theta}^{(j)}\right) \right]^{-1} \Grad \bv{F}\left(\bm{\theta}^{(j)}\right), \quad j \ge 0,
\end{eqnarray}
are performed. Here $\bv{H}$ denotes the Hessian, $\Grad$ denotes the gradient, and the superscript $(j)$ denotes the iteration index. 
In this work the interior-point method was combined with an L-BFGS \cite{liu1989limited} Hessian approximation. Details on the gradients and other aspects of the optimization can be found in Supplementary Information, Section B, at the end of this text. Several stopping criteria were tested, but fixing the number of iterations to 120 provided the most robust reconstructions in all experiments. The regularization parameter $\lambda$ was optimized by grid-search for all experiments.

\subsubsection{Inversion of the reconstructed motion-fields}
\label{section:invertdvf}
Once the optimization algorithm has converged we have representation coefficients of the motion-fields $\bv{T}(\bv{r})$. These are, however, the right inverse motion-fields of $\bv{U}(\bv{r})$ (see \eqref{eq:definitioninversedvf}) which warp the reference image $q_0$ to the dynamic object through \eqref{eq:mainsubstitutionq}. Following \cite{chen2008simple}, the right inverse property of $\bv{T}$ can be rewritten as a relation between the displacements:
\begin{alignat*}{2}
& & \bv{r} &= \bv{U}(\bv{T}(\bv{r})) \\
& & &= \bv{T}(\bv{r}) + \bm{\delta}(\bv{T}(\bv{r})) \\
& & &= \bv{r} + \bm{\eta}(\bv{r}) + \bm{\delta}(\bv{r} + \bm{\eta}(\bv{r})) \\
& \Rightarrow \quad & \bm{\eta}(\bv{r}) &= - \bm{\delta}(\bv{r}+\bm{\eta}(\bv{r})).
\end{alignat*}
We follow \cite{chen2008simple} again and perform fixed-point iterations to compute $\bm{\eta}$:
\begin{alignat}{2} \bm{\eta}^{(0)}(\bv{r}) &= 0, \label{eq:picarditerations1} \\
    \bm{\eta}^{(j)}(\bv{r}) &= -\bm{\delta}\left(\bv{r} + \bm{\eta}^{(j-1)}(\bv{r})\right), \quad j \in \mathbb{N}.
    \label{eq:picarditerations2}
\end{alignat}
Here the superscript $(j)$ denotes the fixed-point iteration index. Note that one iteration results in the naive inversion $\bm{\eta}(\bv{r})=-\bm{\delta}(\bv{r})$, which will only be reasonable for very small deformations. Iterations \eqref{eq:picarditerations1}-\eqref{eq:picarditerations2} are performed until convergence, which takes about 5-10 iterations in practice.

%%%%%%%%------------------------------------------------------
%%%%%%%%------------------------------------------------------

\subsection{Motion estimation experiments}
\label{section:methodsmotionestimation}
\subsubsection{In-silico motion estimation}
In order to validate the proposed MR-MOTUS framework, motion-fields were reconstructed with snapshot data generated from a deforming analytical phantom as described in \autoref{section:modelvalidation}. The signal of the deforming phantom was calculated using \eqref{eq:mrsignal2} with a 3D spiral trajectory (see \autoref{fig:signalcomparison}). To show the potential of the framework to reconstruct motion from highly undersampled $k$-space data, the spiral trajectory consisted of only a total of 78 measured points. Complex noise with mean 0 and standard deviation $2.5 \cdot 10^{-3} n$ was added to the signal. Here $n$ denotes the norm of the simulated $k$-space data. This resulted in a total of 78 noisy signal samples. Affine transformation parameters were reconstructed from the noisy signal by solving \eqref{eq:inverseproblem}. No regularization was applied ($\lambda=0$) and box constraints were added on all parameters to speed up the convergence of the optimization algorithm: for the affine matrix entries we set $|\bv{A}_{ij}|< 1.5$ and for the shift vector elements $|\bv{v}_i|<0.5$. For validation, the reference object was deformed with the reconstructed motion-fields using \eqref{eq:mainsubstitutionq} and compared with the ground-truth object after deformation.
\subsubsection{In-vivo motion estimation}
\label{section:methodsinvivomotionestimation}
To show the practical feasibility of the framework, motion-fields were reconstructed from {\it in-vivo} head (rigid) and {\it in-vivo} abdomen data (non-rigid). In order to have a ground-truth at hand to assess reconstruction quality, a healthy volunteer was scanned with a fully-sampled 3D Cartesian acquisition at several different motion states and the snapshot $k$-space data were retrospectively undersampled from these fully-sampled images. Informed consent was obtained prior to the scans. To reduce any possible effect of motion on the reconstruction quality of the ground-truth Cartesian acquisitions, the scans were acquired in separate breath-holds. 

A 3D spoiled gradient echo sequence was employed, preceded by 200 dummy pulses to reach the required steady-state transverse magnetization. Sequence parameters and other details on both {\it in-vivo} experiments can be found in \autoref{tabel:experimentdetails}. The multi-channel raw $k$-space data were transformed to image space and combined into a single channel to factor out possible influence of coil combinations, and the resulting single-channel images were transformed back to Fourier domain. We will refer to this resulting $k$-space data as the {\it fully-sampled $k$-spaces}, and to the single-channel reconstructions as the {\it fully-sampled reconstructions}. The fully-sampled $k$-spaces were retrospectively undersampled in Fourier domain by truncating the $k$-space. We will refer to this undersampled data as the {\it snapshot $k$-spaces}. This type of undersampling is by no means optimal in terms of the quality of the reconstructed motion-fields, but serves as a proof of concept. Other undersampling patterns are beyond the scope of this work and will be subject of future research. The fully-sampled reconstruction from one motion state was set as reference image, and motion-fields were reconstructed from the retrospectively generated snapshot $k$-spaces at all other motion states. Several undersampling factors were applied to investigate the effect of the amount of available snapshot data on the reconstructed motion-fields. To assess the performance of our framework the reconstructed motion-fields were used to warp the reference image, and the resulting images were compared with the ground-truth fully-sampled reconstructions by calculating the relative differences. As a quantitative comparison, the same procedure was performed for motion-fields obtained by applying optical flow image-registration software \cite{zachiu2015framework,zachiu2015improved} on the fully-sampled reconstructions. 

We have empirically observed that downsampling the reference image ($q_0$) hardly changed the reconstruction quality, but significantly decreases computation time and memory requirements in the optimization. For this reason the reference images were downsampled in $k$-space by a factor of two in all directions in a manner similar to the snapshot $k$-spaces, and then transformed back to image space by a zero-filling reconstruction. The full-resolution reference images were still used to assess the quality of the reconstructed motion-fields, and the twice downsampled reference images were used in the reconstruction algorithm. 

\begin{table*}[tbp]
    \centering
    \caption{In-vivo experiment settings}
    \label{tabel:experimentdetails}
    \begin{tabular}{llll}
        \\
        \hline
         {\bf Parameter} & {\bf In-vivo head experiment} & {\bf In-vivo abdomen experiment}  \\ \hline \hline
         \rowcolor{lightgray} FOV [\si{m}] & $0.25\times0.25\times0.13$ & $0.28\times 0.34\times0.34$ \\ 
         Acquisition size  & $144\times144\times74$ & $94\times128\times128$ \\ 
         \rowcolor{lightgray} Spatial resolution [\si{mm}] & $1.74\times1.74\times1.81$ & $3.00\times2.70\times2.70$ \\  
        
        Repetition time [\si{ms}] & 8.0  & 2.3  \\ 
        \rowcolor{lightgray}Echo time [\si{ms}] & 3.0  &1.2  \\
        Flip angle [\textdegree] & 16 & 20 \\
        \rowcolor{lightgray}Trajectory & Cartesian & Cartesian \\  
        Dummy pulses & 200 & 200 \\
        \rowcolor{lightgray}Pulse sequence type & 3D spoiled GRE & 3D spoiled GRE \\
        Scanner & Philips Ingenia 1.5T & Philips Ingenia 1.5T \\
        \rowcolor{lightgray} Snapshot undersampling & 8, 66, 474, 2551 & 8, 63, 501 \\ 
        Acquired motion states & 7 & 4
        \\
        \rowcolor{lightgray} Motion model (\# parameters) & Affine (12) & Cubic B-Splines (12288)\\
        \hline
    \end{tabular}
    
\end{table*}

%%%%%%%%------------------------------------------------------
%%%%%%%%------------------------------------------------------

\section{Results}
\label{section:results}
\subsection{In-silico motion estimation}
\label{section:insilicomotionestimationresults}
\begin{figure}[tbp]
    \centering
    \includegraphics[width=0.48\textwidth,height=0.52\textwidth]{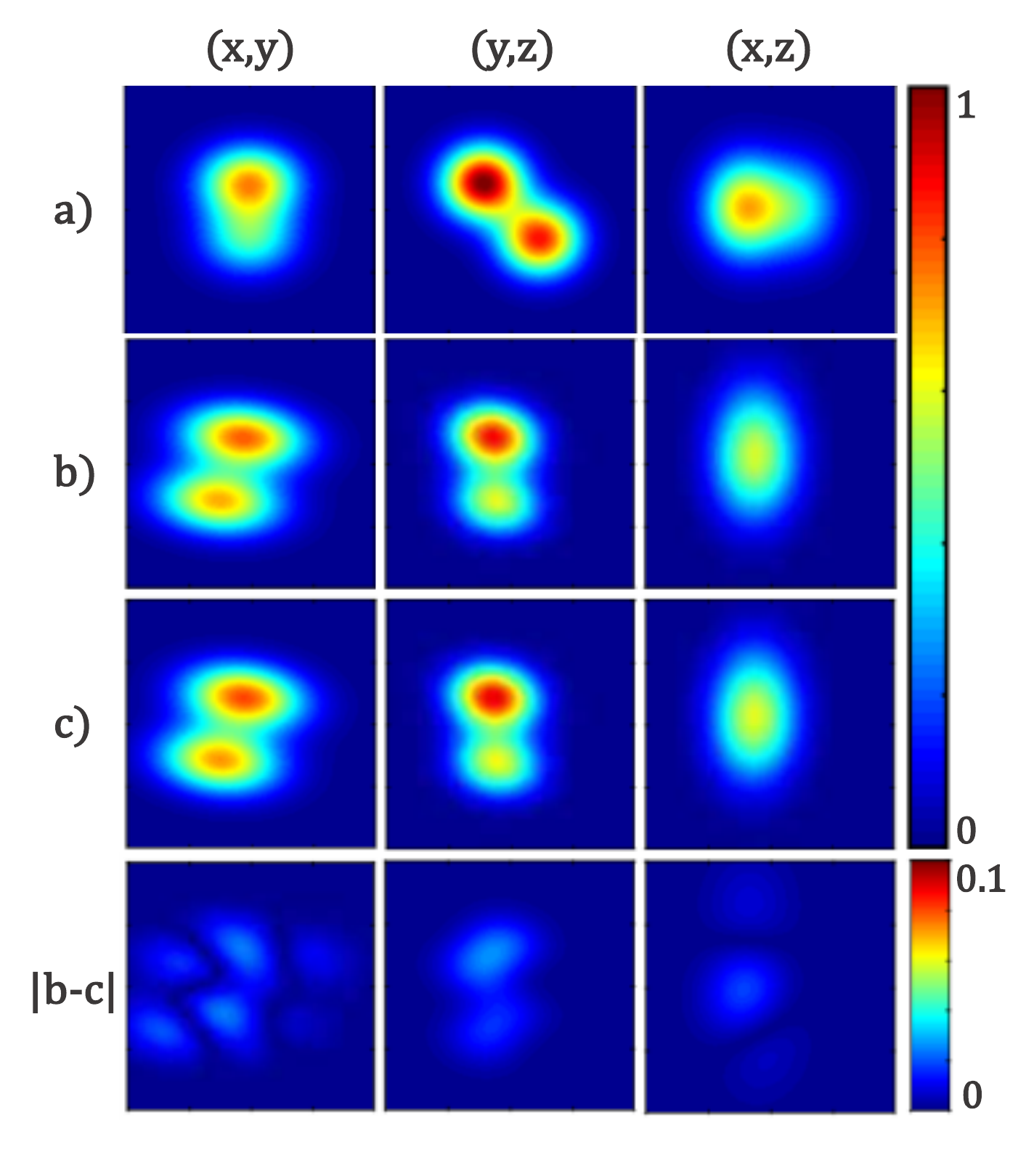}\hfil
\caption{Results from 3D motion reconstruction from the deforming analytical phantom in \autoref{section:insilicomotionestimationresults} using MR-MOTUS: a) shows the object before deformation, b) the ground-truth object after deformation, c) the reconstructed object after deformation, and d) the difference between reconstruction and ground-truth. The image values are in a.u. The reconstructed object in the third row is obtained by calculating \eqref{eq:acspinconservationphantom} with the reconstructed motion-fields. Note that the range for the last row is one order of magnitude smaller than the others. The difference (last row) between the ground-truth (second row) and reconstruction (third row) is minimal, and the small deviations are likely caused by the added noise.}
\label{fig:resultsmodelvalidation}
\end{figure}

The results on in-silico motion reconstructions are presented in \autoref{fig:resultsmodelvalidation}. The reconstructions show that as little as 78 $k$-space points were sufficient to estimate 3D affine motion with high accuracy. The small deviations could be caused by the artificially added complex noise in the simulated data or by interpolation errors.
\subsection{In-vivo motion estimation}
\label{section:invivomotionestimation}
\subsubsection{In-vivo head motion estimation}
\label{section:invivobrain}
\begin{table*}[tbp]
\renewcommand{\arraystretch}{1.3}
\caption{3D rigid head motion estimation}
    \centering
    \begin{tabular}{llll}
        \\
        \hline
        {\bf Method} & {\bf Snapshot undersampling} & {\bf Total \# of $k$-space points} & {\bf Relative error [\%]} \\ \hline \hline
         Optical flow & - & - & 10.30  \\ \hline
         \rowcolor{lightgray} MR-MOTUS  & - & 1534464 & 9.35  \\
         MR-MOTUS & $2.0\times2.0\times2.1=8$ & 186624 & 9.36 \\
        MR-MOTUS & $4.0\times4.0\times4.1=66$&  23328 & 9.59  \\
        \rowcolor{lightgray} MR-MOTUS & $8.0\times8.0\times7.4=474$ & 3240  & 10.77 \\
        MR-MOTUS & $14.4\times14.4\times12.3=2551$ & 600 & 14.79 \\ \hline
    \end{tabular}
\label{table:brainresults}  
\end{table*}
As a first {\it in-vivo} test we applied the MR-MOTUS framework to estimate rigid head motion. The volunteer moved the head along an "8"-shaped trajectory in between a total of seven scans, and held still during actual acquisitions. A dynamic image sequence was obtained by warping the reference image with the reconstructed motion-fields from all motion states. The warped images are compared with the ground-truth fully-sampled reconstructions in the supplementary\footnote{All supplementary Animated Figures are available online at \url{https://surfdrive.surf.nl/files/index.php/s/GCEtiYBxyOnzxjv} and at the arXiv page. A complete description and overview of the supplementary figures can be found in the Appendix and in the README file online.} Animated Figures 1, 2 and 3. The quantitative comparison with optical flow is shown in \autoref{table:brainresults}. The reconstruction times were in the order of minutes for all experiments. 

Good agreement with the ground-truth reconstructions is obtained, even for high undersampling factors up to $474$. For extremely high undersampling factors, when less than 600 $k$-space points are used as input, the recovered motion shows shearing which is not realistic for this head motion. This may, however, be overcome by incorporating a rigidity regularization but this is beyond the scope of this work. The comparison with optical flow shows that similar quality motion-fields can be obtained with MR-MOTUS, while undersampling the input snapshots with factors up to $474$.

\subsubsection{In-vivo respiratory motion estimation}
\label{section:respiratorymotion}
As a second {\it in-vivo} test we applied MR-MOTUS to estimate respiratory motion. We acquired a total of four $k$-spaces at different states in the respiratory cycle. The subject was instructed to move to a different respiratory state in between the acquisitions, and to hold the breath during actual acquisition. 

%%%%%
The reconstruction times were in the order of minutes for all experiments. Similar to what was done for the {\it in-vivo} head motion, the reference image was warped with the reconstructed motion-fields to obtain a dynamic sequence. This dynamic sequence is compared with the fully-sampled ground-truth reconstructions in the supplementary$^2$ Animated Figures 4 and 5. The motion-fields are quantitatively compared with optical flow in \autoref{table:abdomenresults}.

Good agreement is obtained for undersampling factors up to 63, and the relative difference stays within about 2\% of optical flow. For higher undersampling factors the relative error increases rapidly. Visual inspection of reconstructions with extreme undersampling factors shows reasonable deformations in the largest parts of the image, but unrealistic movement in, for example, the spine.
\begin{table*}[tbp]
    \title{\bf Abdomen motion estimation results}
    \centering
    \caption{3D non-rigid respiratory motion estimation.}
    \begin{tabular}{llll}
        \\
        \hline
        {\bf Method} & {\bf Snapshot undersampling} & {\bf Total \# of $k$-space points} & {\bf Relative error [\%]} \\ \hline \hline
        Optical flow & - & - & 31.12  \\ \hline
        \rowcolor{lightgray} MR-MOTUS  & - & 1540096 & 28.79  \\
        MR-MOTUS & $2.0\times2.0\times2.0=8$ & 196608 & 30.24 \\
        \rowcolor{lightgray} MR-MOTUS & $3.9\times4.0\times4.0=63 $ & 24576 & 33.22 \\
        MR-MOTUS & $7.8\times8.0\times8.0=501$ & 3072 & 39.48 \\ \hline
    \end{tabular}
\label{table:abdomenresults} 
\end{table*}
\section{Discussion}
\label{section:discussion}
In this work we introduced MR-MOTUS: a framework for model-based reconstructions of motion-fields from undersampled signals. This framework leverages on the low-dimensionality of physiological motion to reconstruct non-rigid 3D motion-fields from minimal $k$-space data. Direct reconstruction of motion-fields, without the requirement for image reconstruction, can significantly increase the achievable level of acceleration. To this extent, we derived a signal model that explicitly relates $k$-space data to general, possibly non-rigid and/or non-linear motion-fields, by assuming conservation of mass during deformation and a steady-state regime of the transverse magnetization. We hypothesized that motion-fields are inherently low-dimensional, therefore making reconstructions from minimal $k$-space data possible. This hypothesis is supported by our findings: {\it in-vivo} rigid head motion was reconstructed from as little as 600 $k$-space points with a quality comparable with optical flow, and {\it in-vivo} non-rigid 3D respiratory motion was reconstructed with undersampling factors of up to 63 while remaining within about 2\% relative error from state-of-the-art optical flow results. These results show the potential of MR-MOTUS: when combined with a fast and dedicated non-Cartesian trajectory to acquire the highly undersampled $k$-space data, this results in high frame rate 3D non-rigid motion-fields that can open up windows for new applications in cardiac imaging and MR-guided interventions and radiotherapy.

Previously proposed methods to estimate motion directly from $k$-space are mainly based on the Fourier similarity theorems, and therefore limited to affine motion \cite{fu1995orbital,van2006real,welch2002spherical,pipe1999motion}, or based on a joint image and motion reconstruction \cite{burger2018variational,zhao2018coupling,odille2008generalized} such that the achievable level of acceleration is limited by the required image reconstruction. In this work, motion-fields are reconstructed by inverting a signal model. Since the signal model explicitly relates general motion-fields to $k$-space data, the reconstructions are not limited to affine motion, and can be performed directly on $k$-space data without the requirement for image reconstruction. The derived signal model is most closely related to the ones in \cite{burger2018variational,zhao2018coupling}, but the difference is that the joint image/motion reconstruction is transformed into a motion-only reconstruction. This is done by formulating the data-demanding image reconstruction problem explicitly as a motion reconstruction problem through \eqref{eq:mainsubstitutionq2}, thereby circumventing the need for image reconstruction.

The quantitative results presented in this work are promising, but quality is quickly degrading with the amount of undersampling. This is most noticeable at the boundaries of the FOV and in areas where sliding motion between organs occurs. From empirical observations like these, it is clear that significant improvements can still be made in the reconstruction quality by employing e.g. other undersampling patterns, regularization, motion models, or by using multi-channel information in the reconstruction process.

Although the Cartesian undersampling technique used in this work has shown promising results, improvements can be made to capture more motion information and increase the scan efficiency. Since the amount of $k$-space data provided to the algorithm is minimal, it is critical for the reconstruction quality of the motion-fields that this data captures as much information about the motion as possible. The faster the most informative snapshot data can be acquired, the higher the frame-rate and quality in the reconstructed motion-fields. To exemplify this, if the 600 $k$-space points required to reconstruct 3D rigid head motion can be acquired in 60\si{ms}, then the reconstructed 3D motion-fields will have a temporal resolution of 16 frames per second. Future research will therefore address the optimal experimental design of a dedicated, non-Cartesian trajectory that can prospectively and very rapidly acquire the most informative snapshot $k$-space data.

Besides the undersampling technique, further improvements can be made in the choice of regularization and motion model. The curvature regularization and cubic B-spline motion model both penalize discontinuities in the motion-fields, which are present in reality along sliding organ surfaces such as the lung boundaries. Additionally, this choice of motion model forces the resolution to be determined on forehand and it is also not able to exploit the clearly present temporal correlation of motion-fields. In future work we plan to experiment with spatially adaptive, learned and anatomically plausible regularizers (see e.g. \cite{zachiu2018anatomically}). We will also consider multi-resolution and spatio-temporal motion models that may result in even sparser representations. With these extensions the proposed method could be used to reconstruct spatio-temporal non-rigid motion-fields from highly undersampled, prospectively acquired, $k$-space data.

To factor out any possible influence of coil combinations on the reconstruction results we used the Fourier transform of single-channel reconstructions as input data for our reconstructions. However, even higher undersampling factors may be achieved when the signal model is extended to a parallel imaging setting by including the coil sensitivities in the signal model.

With the applications of MR-guided radiotherapy/interventions and cardiac imaging in mind, both the acquisition and reconstruction steps are required to be performed in real-time. In its current state MR-MOTUS estimates 3D motion-fields in about only 2 minutes. This reconstruction time can be reduced even more since hardly any effort has been put into optimization of the reconstruction algorithm. We have used the implementation from \cite{barnett2018parallel} to compute the required type 3 NUFFTs, which scales with the resolution of the reference image and the maximum $k$-space coordinate used to acquire the snapshot data. For this reason, the reference image was downsampled by a factor of two, which resulted in a significant decrease in computation time and memory requirement without degrading the reconstruction quality. In practice, the downsampling also allows a faster acquisition of the reference image. Despite the downsampling, about 80\% of the reconstruction time is devoted to performing type 3 NUFFT computations, and therefore forms the bottleneck in computation speed. Although the implementation from \cite{barnett2018parallel} supports parallel computing on the CPU, a GPU implementation is not yet available. Based on experience, we expect that an acceleration of $10-50$ times in the reconstruction is possible by optimizing the algorithm for speed and by using a GPU implementation of the type 3 NUFFT. This will make the application to online motion reconstruction feasible.

In the derivation of the MR-MOTUS signal model three main assumptions were made: conservation of mass, (no in and outflow of spins), a steady-state of the transverse magnetization, and the availability of an artefact-free reference image. The assumption of mass conservation is almost perfectly valid for head motion but can be invalid in particular regions of the abdomen during respiratory motion. We empirically observed that a violation of the mass conservation property mainly affects the boundaries of the FOV, and hardly degrades the quality of the motion-fields in the middle part of the FOV. The assumption of steady-state magnetization may be partly invalid due to $B_0$ and $B_1$ inhomogeneity or temporal $B_0$ drift. The $B_0$ inhomogeneity and temporal drift can be caused by e.g. hardware imperfections, and may change the equilibrium magnetization - and thereby the steady-state - of moving spins. However, this effect is assumed to be negligible for the combination of small spin displacements and field strengths of up to 1.5T. The $B_1$ inhomogeneity is assumed to be negligible, since single-channel images were used with nearly homogeneous sensitivities. Finally, the availability of an artefact-free reference image can be guaranteed by either employing a more time-consuming acquisition with e.g. gating, since the acquisition of the reference image does not need to be performed in real-time, or by lowering the resolution. We empirically observed that lowering the resolution of the reference image ($q_0$) hardly changed the reconstruction quality of the motion-fields. It will, however, reduce the acquisition time of the reference image, and thereby reduce possible motion artefacts. Lowering the resolution of the reference image has the additional benefit of decreased computation time and memory requirements in the optimization.

Regardless of the possible points of improvement mentioned above, the proposed MR-MOTUS framework has shown promising results to reconstruct non-rigid motion from highly undersampled $k$-space data. We expect that further extensions of will improve upon these results.

\section{Conclusion}
\label{section:conclusion}
In conclusion, we have introduced MR-MOTUS: a framework for model-based reconstruction of motion-fields from undersampled signals. We have derived a dynamic MR-signal model that explicitly relates general motion-fields to $k$-space data. Inversion of this signal model allows to reconstruct non-rigid, non-linear, 3D motion-fields directly from $k$-space. By leveraging on the compressibility of motion-fields, only minimal $k$-space data is required for these reconstruction. This minimal $k$-space data could potentially be acquired within a single TR by employing a fast and dedicated non-Cartesian trajectory, and could therefore result in high frame rate 3D motion-fields that can open up new windows for applications in cardiac imaging, MR-guided interventions and MR-guided radiotherapy.
\section{Acknowledgements}
The authors would like to thank Tom Bruijnen and Oscar van der Heide for the fruitful discussions, and Tom Bruijnen for proofreading the manuscript.

\section*{Appendix}
\section*{Description of supplementary figures}
\noindent This section describes the supplementary Animated Figures referred to in the text. All Animated Figures are available online at \url{https://surfdrive.surf.nl/files/index.php/s/GCEtiYBxyOnzxjv} and the arXiv page.
\newline\newline
\noindent{\bfseries \color{subsectioncolor} AnimatedFigure1.gif}\newline
Affine motion reconstruction from retrospectively undersampled {\it in-vivo} head data using MR-MOTUS, compared with the ground-truth image reconstructions. An undersampling factor of 2.0 was applied in AP and LR direction and 2.1 in FH. This resulted in a total number of 186624 $k$-space points per snapshot and total undersampling factor of about 8.2.\newline \newline
{\bfseries \color{subsectioncolor} AnimatedFigure2.gif}\newline
Affine motion reconstruction from retrospectively undersampled {\it in-vivo} head data using MR-MOTUS, compared with the ground-truth image reconstructions. An undersampling factor of 8.0 was applied in AP and LR directions and 7.4 in FH. This resulted in a total number of 3240 $k$-space points per snapshot and total undersampling factor of about 474.\newline \newline
{\bfseries \color{subsectioncolor} AnimatedFigure3.gif}\newline
Affine motion reconstruction from retrospectively undersampled {\it in-vivo} head data using MR-MOTUS, compared with the ground-truth image reconstructions. An undersampling factor of 14.4 was applied in AP and LR directions and 12.3 in FH. This resulted in a total number of 600 $k$-space points per snapshot and total undersampling factor of 2551. \newline \newline 
{\bfseries \color{subsectioncolor} AnimatedFigure4.gif}\newline
Non-rigid motion reconstruction from {\it in-vivo} abdomen data using MR-MOTUS with a cubic B-spline motion model, compared with ground-truth image reconstructions acquired during breath-hold. No undersampling was applied on the snapshots. \newline \newline
{\bfseries \color{subsectioncolor} AnimatedFigure5.gif}\newline
Non-rigid motion reconstruction from retrospectively undersampled {\it in-vivo} abdomen data using MR-MOTUS with a cubic B-spline motion model, compared with ground-truth image reconstructions acquired during breath-hold. An undersampling factor of about 4 was applied in all direction, resulting in a total undersampling factor of 63. 
\clearpage
\section{Supplementary Information}
\subsection{Detailed MR-MOTUS signal model derivation}
\label{si:detailedmodelderivation}
\begin{figure*}[tbp]
\centering
\includegraphics[width=0.8\textwidth]{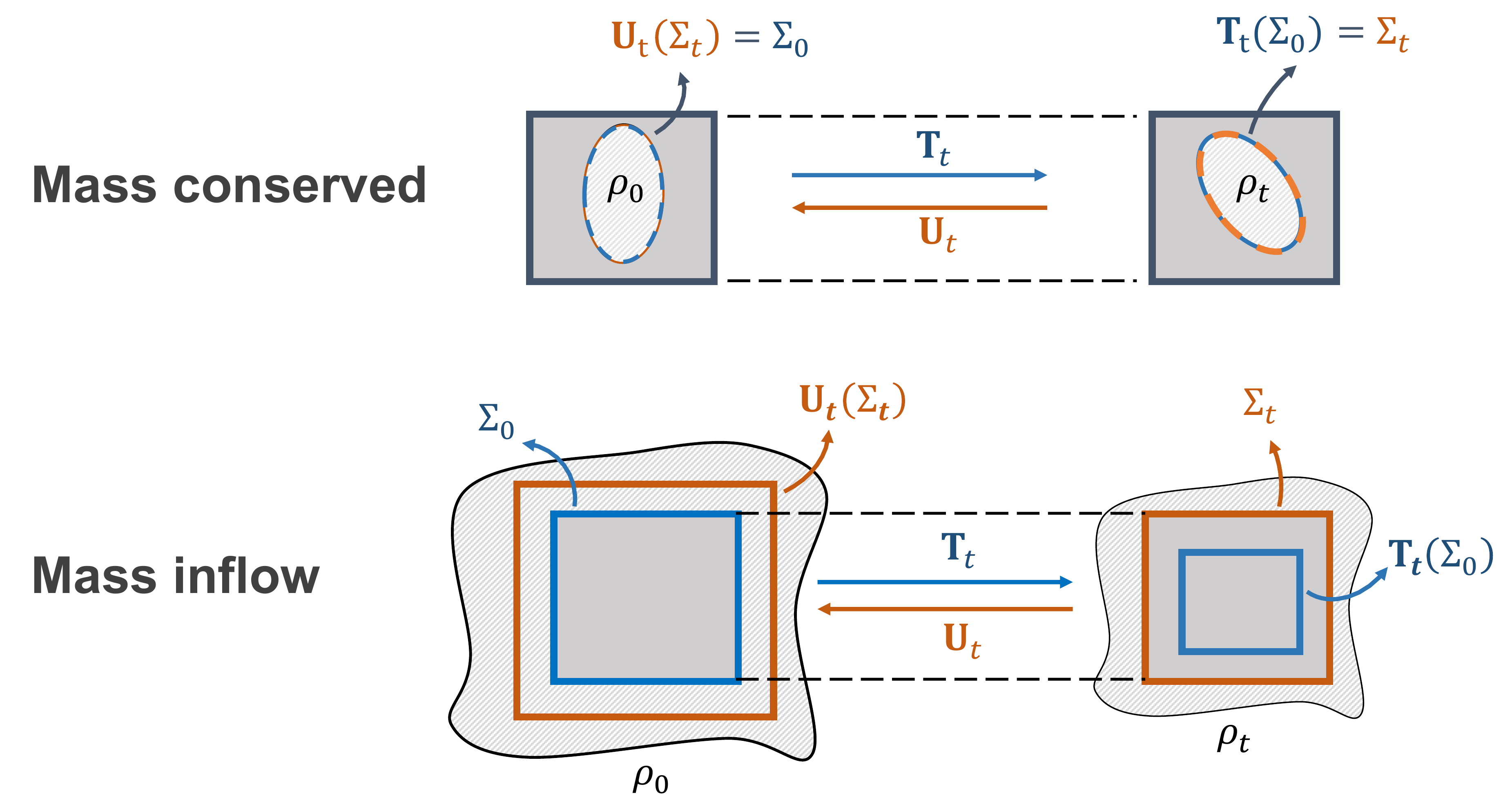}
\caption{{\it This figure is best viewed in color.} Illustration of objects (light-gray) undergoing deformation using the notation introduced in \autoref{si:detailedmodelderivation}: (top) a mass-conserving deformation and (bottom) a deformation with in-flowing mass. The dark-gray squares denote the excited regions at time 0 (left) and time $t$ (right). $\bv{T}_t(\Sigma_0)$ denotes the new locations after deformation at time $t$, of spins in $\Sigma_0$ that were originally excited for the reference image. In this area the signal can be described with the model from Equation (14) in the main text. $\bv{U}_t(\Sigma_t)$ are the source locations of the spins that are excited for the $t$-th snapshot image. If this area does not coincide with $\Sigma_0$, then the signal from the area between $\Sigma_0$ and $\bv{U}_t(\Sigma_t)$ cannot be explained by the model. A similar observation can be made for deformations with mass out-flow, except that in that case $\bv{U}_t(\Sigma_t)$ will lie within $\Sigma_0$.
}
\label{fig:spaces}
\end{figure*}

Under the assumptions made in the main text (see Section II-B) the steady-state equilibrium of all individual spins remains nearly constant during motion. Hence, the transverse magnetization at time $t$ of the spins at location $\bv{r}_t$ can be written as the transverse magnetization at time $0$ of the same spin before deformation by $\bv{U}_t$. Defining the support of $m_t$ as $\Sigma_j$, the observation above can mathematically be summarized as (see \autoref{fig:spaces})
 \begin{equation}
     m_t(\bv{r}_t) = m_0\left(\bv{U}_t(\bv{r}_t)\right), \ \ \bv{r}_t \in \bv{T}_t(\Sigma_0) \ \cap \ \Sigma_t. \label{eq:steadystatederiv1_appendix}
 \end{equation}
From the assumption on the conservation of mass it was derived in the main text that the following must hold for all $t=0,1,\dots$ and $\bv{r} \in \mathbb{R}^3$:
\begin{equation}
    \label{eq:acspinconservation_appendix}
    \rho_t(\bv{r}_t)\infd \bv{r}_t = \rho_0\left(\bv{U}_t(\bv{r}_t)\right)\lvert \det(\Grad \bv{U}_t ) (\bv{r}_t)\rvert \infd \bv{r}_t.
\end{equation}
Combining \eqref{eq:steadystatederiv1_appendix} and \eqref{eq:acspinconservation_appendix} into one equation , with $q_t \equiv \rho_t \cdot m_t$, yields, 
\begin{equation}
    \label{eq:mainsubstitutionq2_appendix}
    q_t(\bv{r}_t)\infd \bv{r}_t= q_0\left(\bv{U}_t(\bv{r}_t)\right)\lvert \det(\Grad \bv{U}_t) (\bv{r}_t)\rvert \infd \bv{r}_t,
\end{equation}
for $\bv{r}_t \in \bv{T}_t(\Sigma_0) \cap \Sigma_t$.
The signal at time $t$ is given as
\begin{equation}
s_t(\bv{k}) = \int_{\Sigma_t} q_t(\bv{r}_t) e^{-i 2 \pi \bv{k} \cdot \bv{r}_t} \ \infd \bv{r}_t,  \label{eq:signalnormal_appendix}
\end{equation}
hence the substitution of \eqref{eq:mainsubstitutionq2_appendix} into \eqref{eq:signalnormal_appendix} is only valid when $\Sigma_t \subseteq \bv{T}_t(\Sigma_0)$ holds. In other words, the substitution is valid when there is no in-flow of new spins between time 0 and time $t$. Under that assumption we get
\begin{eqnarray*}
s_t(\bv{k}) &=& \int_{\Sigma_t} q_0\left(\bv{U}_t(\bv{r}_t)\right) e^{-i 2 \pi \bv{k} \cdot \bv{r}_t} \ \lvert \det(\Grad \bv{U}_t ) (\bv{r}_t) \rvert \ \ \infd \bv{r}_t,
\end{eqnarray*}
and after the change of variables $\bv{r}_t \mapsto \bv{T}_t(\bv{r}_0)$ we obtain the general signal model
\begin{equation}
\label{eq:signalmodel_appendix}
s_t(\bv{k}) = \int_{\bv{U}_t(\Sigma_t)} q_0(\bv{r}_0) e^{-i 2 \pi \bv{k} \cdot \bv{T}_t(\bv{r}_0)} \ \infd \bv{r}_0.
\end{equation}
Note that the domain of integration in \eqref{eq:signalmodel_appendix} depends on the (unknown) motion field $\bv{U}_t$, which is inconvenient in practice. If $\Sigma_0 \subseteq \bv{U}_t(\Sigma_t)$, i.e. the support of $q_0$ is contained in $\bv{U}_t(\Sigma_t)$, then the integration limit can be changed to $\Sigma_0$. However, since the substitution of \eqref{eq:mainsubstitutionq2_appendix} into \eqref{eq:signalnormal_appendix} already assumes $\Sigma_t \subseteq \bv{T}_t(\Sigma_0)$, this implies that $\Sigma_0 = \bv{U}_t(\Sigma_t)$ must hold. In other words, the general model \eqref{eq:signalmodel_appendix} and the MR-MOTUS signal model in Equation (14) in the main text are equivalent when always the exact same spins are excited for the reference and all subsequent snapshots (see \autoref{fig:spaces} for an illustration). This happens when no mass flows in or out of the FOV between the acquisition of the reference image and subsequent snapshot acquistions.

\subsection{Reconstruction implementation details}
\label{si:implementationdetails}
\subsubsection{Motion models}
\label{section:motionmodels}
The motion models are implemented in matrix form as 
\[\bv{R}_t = \bv{R}_0 + \bv{B}\bm{\theta}_t = \begin{pmatrix} \bv{R}_0^x \\ \bv{R}_0^y \\ \bv{R}_0^z \end{pmatrix} + \begin{pmatrix}\bv{B}^x \\ & \bv{B}^y \\ & & \bv{B}^z \end{pmatrix} \begin{pmatrix}\bm{\theta}_t^x \\ \bm{\theta}_t^y \\
\bm{\theta}_t^z \end{pmatrix}.\] Here $\bv{R}_0$ and $\bv{R}_t$ denote the vectors with coordinates respectively at time 0 and after deformation by $\bv{T}_t$ at time $t$, $\bv{B}$ is the matrix determined by the motion model, and $\bm{\theta}_t$ is the vectors of unknowns. Both $\bv{R}_0$ and $\bv{R}_t$ are obtained by first concatenating vertically per dimension and then stacking the dimensions vertically. Hence, for a total number of $N_r$ voxels we have $\bv{R}_0,\bv{R}_t\in \mathbb{R}^{3N_r}.$ The affine model introduced in Equation (20) in the main text can be obtained by setting 
\begin{alignat*}{2} &\bv{B}^x = \bv{B}^y = \bv{B}^z = \left[\bv{R}_0^x \ \  \bv{R}_0^y \ \ \bv{R}_0^z \ \ \bv{1} \right], \\ & \begin{pmatrix} \bm{\theta}_t^x \\ \bm{\theta}_t^y \\ \bm{\theta}_t^z \end{pmatrix} = \text{vec}\left\{ \begin{pmatrix} \bv{A}_t^T - \bv{I}  \\ \bv{v}_t^T \end{pmatrix} \right\}, \end{alignat*}
where $\bm{\theta}_t^p \in \mathbb{R}^{4 \times 1}$, $\bv{1} \in \mathbb{R}^{N_r \times 1}$ denotes an all-one vector, and $\text{vec}(\cdot)$ the vectorization function. For the spline model that was introduced in Equation (21) in the main text we get
 $\bv{B}^p \in \mathbb{R}^{N_r \times N_c^p}$ as B-spline basis matrices with vertically concatenated $\bv{b}^p(\bv{r}) \in \mathbb{R}^{1\times N_c^p}$ as entries. The $\bm{\theta}_t^p \in \mathbb{R}^{N_c^p \times 1}$ denote the basis coefficients.

%%%%%%%%------------------------------------------------------
%%%%%%%%-----------------------------------------------------

\subsubsection{Gradient of the objective function}
To compute the gradient with respect to the objective function we first write the forward model in a way that is more convenient for differentiation:
$$ \bv{F}(\bm{\theta}_t \hspace{0.1cm}  | \hspace{0.1cm} \bv{q}_0 ) = \exp \left\{ -i 2\pi \bv{K}\left[\bv{R}_t^x \ \ \bv{R}_t^y \ \ \bv{R}_t^z\right]^T \right\} \bv{q}_0 , $$
where $\bv{K}\in\mathbb{R}^{M \times 3}$ is the matrix with all $k$-space coordinates of the applied trajectory, exp\{$\cdot$\} denotes element-wise application of the exponential function, and $\bv{q}_0$ denotes the vectorization of $q_0$. Similarly, the curvature regularization function can be written as a simple $L^2$-norm of a matrix-vector product by applying a finite difference scheme to discretize the derivatives, i.e.
\[\begin{pmatrix}\bv{\Delta}\bv{T}_t^x \\ \bv{\Delta}\bv{T}_t^y \\ \bv{\Delta}\bv{T}_t^z \end{pmatrix} \approx  \begin{pmatrix}\textbf{\L} \\ & \textbf{\L} \\ & & \textbf{\L} \end{pmatrix}\begin{pmatrix} \bv{R}_0^x + \bv{B}^x \bm{\theta}_t^x \\ \bv{R}_0^y + \bv{B}^y \bm{\theta}_t^y \\ \bv{R}_0^z + \bv{B}^z \bm{\theta}_t^z \end{pmatrix} = \bv{L}(\bv{R}_0+\bv{B}\bm{\theta}_t). \] Here $\Delta \bv{T}^p,\bv{T}^p \in \mathbb{R}^{N_r\times 1}$ are vectorizations of the corresponding functions, and $\textbf{\L}\in\mathbb{R}^{N_r \times N_r}$ is defined as the matrix that applies the dicretized Laplace operator to a vector by left-multiplication. The discretization of the spatial derivatives was obtained using a central difference scheme with Neumann boundary conditions. Using this, we get 
\[\mathcal{R}(\bv{T}_t) = \lVert \bv{L}(\bv{R}_0+\bv{B}\bm{\theta}_t) \rVert_2^2. \]  
The total objective function can now be written in discretized form as
\[E(\bm{\theta}_t)=\lVert \exp \left\{ -i 2\pi \bv{K}\widetilde{\bv{R}}_t^T \right\} \bv{q}_0 - \bv{s} \rVert_2^2 + \lambda \lVert \bv{L}\bv{R}_t \rVert_2^2, \]
where we have defined $\widetilde{\bv{R}}_t:=\left[\bv{R}_t^x \ \bv{R}_t^y \ \bv{R}_t^z  \right]$ as the new coordinates at time $t$ dependent on $\bm{\theta}_t$. The Jacobian $\bv{J}_F$ is defined as
\[ \bv{J}_F := \left[\pdiff{\bv{F}}{\bm{\theta}_t^x},\pdiff{\bv{F}}{\bm{\theta}_t^y},\pdiff{\bv{F}}{\bm{\theta}_t^z}\right] \in \mathbb{C}^{N_k \times N_c}.\]
Using matrix differentiation we can derive the following formulas for the derivatives
\begin{eqnarray}
    \pdiff{\bv{F}}{\bm{\theta}_t^p} =& -i 2\pi \text{diag}(\bv{k}^p) \cdot \dots & \nonumber \\
    & \exp \left\{ -i 2\pi \bv{K}  \widetilde{\bv{R}}_t^T \right\} \cdot \dots & \label{eq:derivatives1} \\ & \text{diag}(\bv{q}_0)\bv{B}^p, & \nonumber \\ \nonumber
    \left(\pdiff{\bv{F}}{\bm{\theta}_t^p}\right)^\star =& i 2\pi \left(\bv{B}^p\right)^T \text{diag}(\bar{\bv{q}}_0) \cdot \dots & \\ & \label{eq:derivatives2} \exp \left\{ i 2\pi  \widetilde{\bv{R}}_t \bv{K}^T  \right\} \cdot \dots & \\ & \text{diag}(\bv{k}^p).\nonumber
\end{eqnarray} 
Note the complex conjugate $\bar{\bv{q}}_0$ and the change in signs of the imaginary variables in \eqref{eq:derivatives2}. The gradient of the objective function is readily computed as
\begin{equation*}
\pdiff{E}{\bm{\theta}_t} = \begin{pmatrix} \partial E / \partial \bm{\theta}_t^x \\ \partial E / \partial \bm{\theta}_t^y\\ \partial E / \partial \bm{\theta}_t^z \end{pmatrix} = 2 \text{Re}\left\{ (\bv{J}_F)^\star(\bv{F}-\bv{s}_t) \right\} + 2 \lambda \bv{B}^T \bv{L}^T \bv{L}\bv{R}_t.
\end{equation*}
Here the superscript star denotes the Hermitian adjoint. 

\subsubsection{Jacobian-vector products}
In the optimization algorithm frequent computations of the products $\bv{J}_F \bv{w}_1$ and $(\bv{J}_F)^\star\bv{w}_2$ are required. By definition 
\begin{alignat*}{3}
\bv{J}_F\bv{w}_1 = \sum_{p\in\{x,y,z\}} \left(\pdiff{\bv{F}}{\bm{\theta}_t^p}\right) \bv{w}_1^p, &\qquad (\bv{J}_F)^\star \bv{w}_2 = \begin{pmatrix} \left(\pdiff{\bv{F}}{\bm{\theta}_t^x}\right)^\star \bv{w}_2 \\ \left(\pdiff{\bv{F}}{\bm{\theta}_t^y}\right)^\star \bv{w}_2 \\ \left(\pdiff{\bv{F}}{\bm{\theta}_t^z}\right)^\star \bv{w}_2 \end{pmatrix}.
\end{alignat*}
Here we have defined $\bv{w}_1^p$ such that $\bv{w}_1:=\begin{pmatrix} \bv{w}_1^x \\ \bv{w}_1^y \\ \bv{w}_1^z \end{pmatrix}$. Using \eqref{eq:derivatives1}-\eqref{eq:derivatives2} we can compute 
\begin{equation*}
    \bv{J}_F \bv{w}_1=-i 2\pi \sum_{p\in\{x,y,z\}}\left\{\right. \text{diag}(\bv{k}^p) \exp \left\{ -i 2\pi \bv{K} \widetilde{\bv{R}}_t^T \right\} \cdot \dots
\end{equation*}
\begin{equation*}
    \hspace{1cm}\text{diag}(\bv{q}) \bv{B}^p \bv{w}_1^p \left.\right\}
\end{equation*}
and
\begin{equation*}
(\bv{J}_F)^\star \bv{w}_2 = i 2\pi \begin{pmatrix} \left(\bv{B}^x\right)^T \text{diag}(\bar{\bv{q}}) \exp \left\{ i 2\pi \widetilde{\bv{R}}_t \bv{K}^T  \right\} \text{diag}(\bv{k}^x) \bv{w}_2 \\ \left(\bv{B}^y\right)^T \text{diag}(\bar{\bv{q}}) \exp \left\{ i 2\pi \widetilde{\bv{R}}_t \bv{K}^T  \right\}  \text{diag}(\bv{k}^y) \bv{w}_2  \\ \left(\bv{B}^z\right)^T  \text{diag}(\bar{\bv{q}}) \exp \left\{ i 2\pi \widetilde{\bv{R}}_t \bv{K}^T  \right\}  \text{diag}(\bv{k}^z) \bv{w}_2 \end{pmatrix}.
\end{equation*}
The first product is implemented as a forward type 3 NUFFT, and the second product is implemented as a a backward type 3 NUFFT. For both we have used the efficient implementations provided by [29].

\end{document}